\documentclass{ws-ijmpa}

\usepackage{amsmath,amssymb}
\usepackage{graphicx}
\usepackage{booktabs}
\usepackage{listings}
\usepackage{xcolor}
\makeatletter
\providecommand{\captionof}[2]{%
  \refstepcounter{#1}%
  \par\noindent\@makecaption{\csname #1name\endcsname~\csname the#1\endcsname}{#2}\par
}
\makeatother
\usepackage{fox-listings}
\usepackage{algorithm}
\usepackage{algorithmic}

\makeatletter
\newcommand\fs@dblruled{%
  \def\@fs@cfont{\bfseries}%
  \let\@fs@capt\floatc@ruled
  \def\@fs@pre{\hrule height0.5pt depth0pt \kern2pt\hrule height0.5pt depth0pt \kern2pt}%
  \def\@fs@post{\nobreak\vskip2pt\hrule height0.5pt depth0pt \nobreak\vskip2pt\hrule height0.5pt depth0pt}%
  \def\@fs@mid{\kern2pt\hrule\kern2pt}%
  \let\@fs@iftopcapt\iftrue}
\makeatother
\floatstyle{dblruled}
\restylefloat{algorithm}

\graphicspath{{figures/}}

\newcommand{\xichromy}{\xi_{y}}
\newcommand{\xichromx}{\xi_{x}}
\newcommand{\Vesq}{V_{\mathrm{ESQ}}}
\newcommand{\xiyana}{\xi_{y}^{(p),\,\mathrm{ana}}}
\newcommand{\xiycosy}{\xi_{y}^{(p),\,\mathrm{COSY}}}

\begin{document}

\markboth{E. Valetov, K. Makino, and M. Berz}{Vertical chromaticity of the Fermilab Muon \textit{g}-2 storage ring}

\catchline{}{}{}{}{}

\title{Analytic Derivation of Vertical Chromaticity in the Fermilab Muon $g{-}2$ Storage Ring\thanks{Fermilab report FERMILAB-PUB-26-0333-PPD.}}

\author{EREMEY VALETOV, KYOKO MAKINO, and MARTIN BERZ}
\address{Department of Physics and Astronomy, Michigan State University,\\
East Lansing, MI 48824, USA\\
Muon $g{-}2$ Collaboration, Fermi National Accelerator Laboratory,\\
Batavia, IL 60510, USA\\
evv@msu.edu (corresponding author), makino@msu.edu, berz@msu.edu}

\maketitle

\begin{history}
\received{}
\revised{}
\end{history}

\begin{abstract}
We derive the vertical chromaticity $\xichromy$ of the Fermilab Muon $g{-}2$
storage ring in closed analytic form. Expanding the Hamiltonian as a
Taylor polynomial in the dynamical variables and integrating the equations
of motion order by order, we obtain the
vertical second-order aberrations of the homogeneous magnetic dipole
($\mathtt{DI}$) and the combined-function dipole-and-electrostatic-quadrupole
element ($\mathtt{DIQ}$) used in the muon $g{-}2$ ring. Composing the per-element
maps over the periodic dispersion orbit yields a closed-form expression for the
vertical chromaticity $\xichromy$ of the continuous-ring
$\mathtt{DIQ360}$ model, in direct functional analogy with the horizontal
result of our earlier work on the same ring (Ref.~\refcite{ChromCPO11}).
Comparison against COSY INFINITY differential-algebra computation shows
agreement at the $10^{-11}$ level across all three ring models ($\mathtt{DIQ360}$
closed form and the modular $\mathtt{DIEQ\_ON}$, $\mathtt{DIEQ}$ via per-element
composition) for muon $g{-}2$ electrostatic-quadrupole (ESQ) voltages
$\Vesq \in [10, 26]\,\mathrm{kV}$.
\end{abstract}

\section{Introduction}
\label{sec:intro}

The Fermilab Muon $g{-}2$ Experiment (E989)\cite{g2TDR} recently reported
the muon anomalous magnetic moment with a precision of
$127\,\mathrm{ppb}$\cite{g2PRL25,g2PRD24}, the most precise measurement of
$a_\mu$ to date. Extracting $a_\mu$ from the measured spin-precession
frequency $\omega_a$ requires beam-dynamics corrections that are sensitive
to the closed orbit, betatron tunes, and chromaticities of the storage
ring\cite{g2PRAB21,weisskopfphd}. Among these, the vertical
chromaticity $\xichromy$ is a fundamental beam-dynamics observable of
the muon storage ring; its analytic closed-form expression is the
subject of the present work.

We previously derived the analytic horizontal second-order aberrations
of the homogeneous magnetic dipole ($\mathtt{DI}$) and the
combined-function dipole-and-electrostatic-quadrupole element
($\mathtt{DIQ}$) of the muon $g{-}2$ ring, together with a closed-form
horizontal chromaticity $\xichromx$\cite{ChromCPO11},
using the order-by-order Hamiltonian perturbation method.
This method, introduced in Karl Brown's foundational
work\cite{Brown:1968hkv,bbb64,Brown1981,Brown:1984qw} and developed in
the Hamiltonian treatments of
Refs.~\refcite{AIEP108book,IP127book}, expresses the equations of
motion as truncated power series: the linear part is solved exactly
(a matrix exponential for constant-coefficient elements), and each
higher order is obtained iteratively by
treating the lower-order solution as a driving term in an aberration
integral. We have also applied it to electrostatic
deflectors in Refs.~\refcite{ELSPHTM17,ESCPO10AIEP}. The vertical
chromaticity has been computed numerically by the
differential-algebra (DA) techniques of
Refs.~\refcite{AIEP108book,COSYCAP04,COSYCPO11} but has lacked an
analytic counterpart. The present work supplies that counterpart: a
closed-form $\xichromy$ for the continuous-ring
$\mathtt{DIQ360}$ model, together with the full vertical second-order
aberration table for $\mathtt{DI}$ and $\mathtt{DIQ}$, validated
numerically against COSY INFINITY DA across a sweep of the
muon $g{-}2$ storage-ring operating voltages.

\section{The Muon $g{-}2$ Storage Ring}
\label{sec:ring}

\begin{figure}[t]
  \centering
  \includegraphics[width=0.85\linewidth]{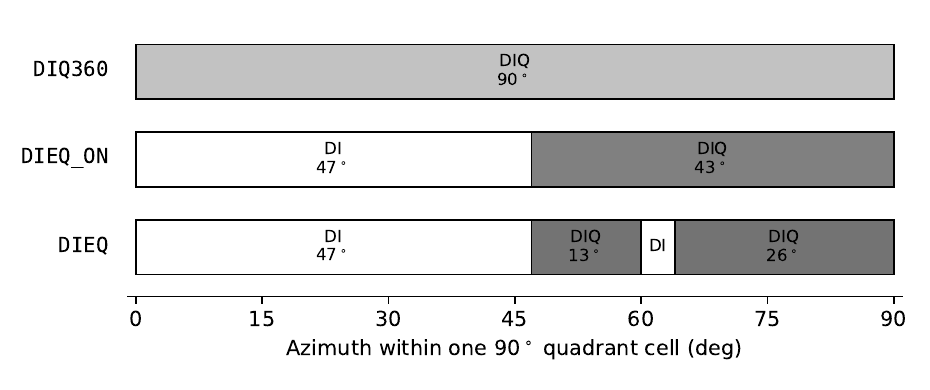}
  \caption{Element layout of the three ring models over one $90^\circ$
    quadrant cell: the combined-function dipole-and-ESQ element
    $\mathtt{DIQ}$ (gray) and the homogeneous magnetic dipole
    $\mathtt{DI}$ (white). $\mathtt{DIQ360}$ is a single $360^\circ$
    $\mathtt{DIQ}$ (here $90^\circ$ over one cell) with the ESQ
    inhomogeneity azimuthally averaged; $\mathtt{DIEQ\_ON}$ lumps the
    quadrant ESQ into one $43^\circ$ $\mathtt{DIQ}$; $\mathtt{DIEQ}$
    resolves the true short/long ESQ split. The $\mathtt{DIQ}$ gray
    level is a schematic normalization, scaled inversely with the
    total $\mathtt{DIQ}$ arc so that the apparent focusing strength
    (gray level $\times$ arc) is visually comparable across the three
    models rather than tracking arc length alone; $\mathtt{DI}$
    sections are white. The residual differences between the models
    are quantified in Sec.~\ref{sec:validation}.
    \label{fig:lattice-models}}
\end{figure}

\begin{figure}[t]
  \centering
  \includegraphics[width=0.825\linewidth]{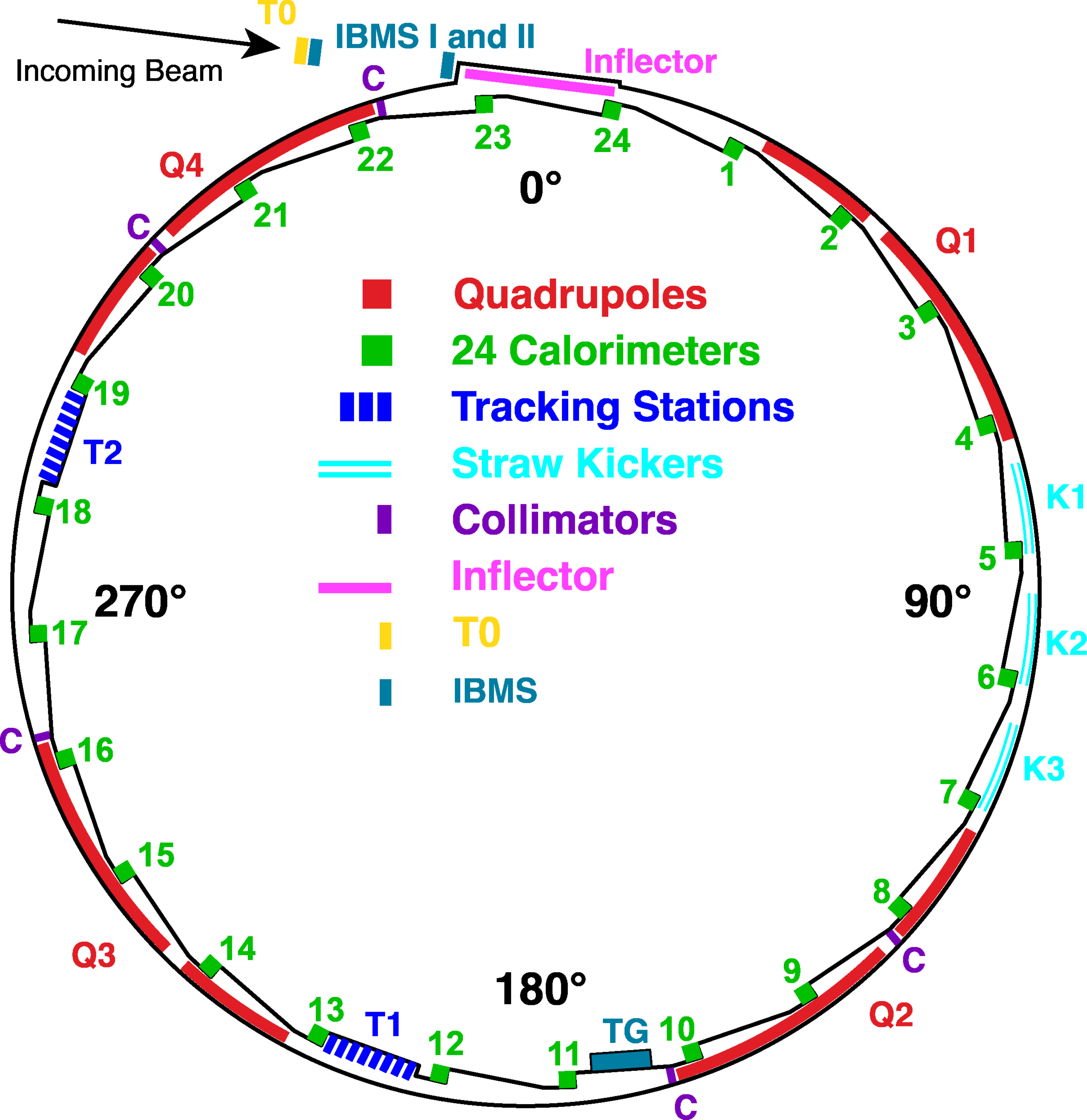}
  \caption{System diagram of the muon $g{-}2$ storage ring. The four
    electrostatic quadrupole stations Q1--Q4 provide vertical focusing; the
    fast and slow muon kickers (K1--K3) inject the beam onto the closed
    orbit. (Adapted from Ref.~\protect\refcite{g2PRAB21}, CC~BY~4.0.)}
  \label{fig:ring-diagram}
\end{figure}

The muon $g{-}2$ storage ring at Fermilab consists of a homogeneous vertical
magnetic-dipole field of $1.45\,\mathrm{T}$ providing a closed circular
reference orbit of radius $R_0 = 7.112\,\mathrm{m}$, and four
electrostatic-quadrupole (ESQ) stations providing vertical focusing. The reference muon momentum
is the nominal magic momentum $p_0 = 3094\,\mathrm{MeV}/c$\cite{g2TDR},
at which the electric-field contribution to the spin-precession
frequency vanishes at first order\cite{g2PRD24}. We use the
reference Lorentz factor\cite{g2LinearOptics17,ChromCPO11}
\begin{equation}
  \gamma_0 = 29.300124824596928.
  \label{eq:gamma0}
\end{equation}
The four ESQ stations together cover approximately $43\%$ of the ring azimuth.
A schematic of the storage ring is shown in Fig.~\ref{fig:ring-diagram}.
Three ring models are studied here. The continuous ring $\mathtt{DIQ360}$ is a
single $360^\circ$ $\mathtt{DIQ}$ element with the four ESQ stations represented
by their azimuthally averaged inhomogeneity; this is the form for which a
closed-form $\xichromy$ is obtained in Sec.~\ref{sec:result}. The
simplified modular ring $\mathtt{DIEQ\_ON}$ lumps the two ESQ
elements of each quadrant into a single $43^\circ$ $\mathtt{DIQ}$
element, giving four cells of $\mathtt{DI}\,47^\circ + \mathtt{DIQ}\,43^\circ$
per quadrant (90$^\circ$). This is the geometry used in
Ref.~\refcite{onkim} for the horizontal chromaticity derivation, and
lies between $\mathtt{DIQ360}$ and the full modular structure. The full modular ring $\mathtt{DIEQ}$ resolves each quadrant
into four elements ($\mathtt{DI}\,47^\circ + \mathtt{DIQ}\,13^\circ +
\mathtt{DI}\,4^\circ + \mathtt{DIQ}\,26^\circ$), reproducing the actual
short-arc / long-arc ESQ split of the ring; the closed-form modular composition
of Sec.~\ref{sec:validation} uses these four-element per-quadrant maps.
The element layout of the three models over one $90^\circ$ quadrant
cell is shown in Fig.~\ref{fig:lattice-models}.

\section{Methodology}
\label{sec:methods}

The Hamiltonian framework for arbitrary-order
charged-particle optical-element aberrations is laid out in
Ref.~\refcite{AIEP108book}; the present formulation parallels its
application to curvilinear electrostatic elements in
Ref.~\refcite{ESCPO10AIEP}. The COSY INFINITY beamline coordinate
system $(x, a, y, b, \ell, \delta_K)$\cite{COSYBeamMan102,COSYCAP04,COSYCPO11}
is used throughout, with $a = p_x/p_0$, $b = p_y/p_0$ the canonical transverse
momenta normalised to the reference momentum $p_0$, and $\delta_K = (K - K_0)/K_0$
the relative kinetic energy deviation. Vertical
chromaticity is taken in the momentum-based convention
\begin{equation}
  \xichromy \;=\; \left[\partial \nu_y/\partial \delta_p\right]_{\delta_p = 0},
  \qquad \delta_p = (p - p_0)/p_0,
  \label{eq:xidef}
\end{equation}
which is the convention used internally by the Muon $g{-}2$ Experiment
analysis chain\cite{g2PRAB21,weisskopfphd}.
All chromaticities below are evaluated for the on-momentum-designed
lattice at the off-momentum dispersive closed orbit (fixed-lattice
convention); the alternative scaling convention, in which the lattice
is rescaled with momentum, gives a numerically and conceptually
different quantity that omits the dispersion-orbit-coupling contribution.
The conversion between $\delta_K$- and $\delta_p$-based chromaticities is
the exact factor $\xichromy^{(p)} = \xichromy^{(K)} \cdot (1+\gamma_0)/\gamma_0$,
which differs from unity by about $3\%$ at the magic momentum.

We adopt the multiplicity-factorial aberration coefficient convention of
Ref.~\refcite{AIEP108book}: for output coordinate $z_i$ and input monomial
$z_{j_1}^{k_1} \cdots z_{j_m}^{k_m}$ with distinct coordinate indices
$j_1, \ldots, j_m$ and multiplicities $k_p \ge 1$,
\begin{equation}
  \left(z_i|z_{j_1}^{k_1} \cdots z_{j_m}^{k_m}\right) \;=\; \frac{1}{k_1!\cdots k_m!}\;
  \frac{\partial^{k_1+\cdots+k_m} \mathcal{M}(\mathbf{z})_i}{\partial z_{j_1}^{k_1}\cdots \partial z_{j_m}^{k_m}}
  \bigg|_{\mathbf{z} = 0}.
  \label{eq:abdef}
\end{equation}
Reference~\refcite{ChromCPO11} derived the horizontal second-order
aberrations of the homogeneous magnetic dipole ($\mathtt{DI}$) and the
combined-function dipole-and-electrostatic-quadrupole element ($\mathtt{DIQ}$)
used in the muon $g{-}2$ ring via order-by-order perturbation of the
canonical equations of motion; the vertical second-order coefficients
were not derived there. The present work supplies them.

The Hamiltonian is expanded as a Taylor polynomial in the dynamical
variables,
\begin{equation}
  H = H_2 + H_3 + \cdots,
  \label{eq:Hexp}
\end{equation}
where the constant term $H_0$ is omitted because it does not enter the
equations of motion, and the first-order term $H_1$ vanishes for an
expansion about the reference orbit. The quadratic part $H_2$ governs
the linear motion; within each element its coefficients are constant, so
the linear map $L$ is obtained by exponentiating the constant linear
generator, and the full-ring map follows by composition of the
per-element maps. The cubic part $H_3$ drives the second-order map
through the aberration integral
\begin{equation}
  R_2(s) \;=\; L(s)\,\int_0^s L^{-1}(s')\,Q_2\bigl(s', \mathbf{z}_{\mathrm{lin}}(s')\bigr)\,ds',
  \label{eq:VOP}
\end{equation}
with $Q_2$ the second-order driving evaluated on the linear trajectory
$\mathbf{z}_{\mathrm{lin}}(s) = L(s)\,\mathbf{z}_0$.
The curvilinear $(1{+}hx)$ factor in $H$ already encodes the
Frenet--Serret rotating reference frame, so no additional Coriolis or
centripetal term is introduced when expanding about the design orbit.

A Wolfram Language implementation of the order-by-order procedure of
Refs.~\refcite{AIEP108book,ESCPO10AIEP} generates the $5{\times}20$
second-order aberration coefficients for $\mathtt{DI}$
and $\mathtt{DIQ}$ in the COSY INFINITY beamline coordinates. Its horizontal
subset reproduces all 27 coefficients of Ref.~\refcite{ChromCPO11} under
\texttt{FullSimplify}; the vertical $(y, b)$-row coefficients are obtained from the same
pipeline. Algorithm~\ref{alg:perturb}
summarises the procedure; the COSY INFINITY DA validation program
is shown in Listing~\ref{lst:sweep}.

\begin{algorithm}[t]
\caption{Order-by-order Hamiltonian aberration computation, as
implemented for the muon $g{-}2$ ring elements. Coefficients are returned
in the multiplicity-factorial convention of Eq.~\ref{eq:abdef}.}
\label{alg:perturb}
\begin{algorithmic}[1]
\REQUIRE Hamiltonian $H(\mathbf{z}; s)$ in curvilinear coordinates
$\mathbf{z} = (x, a, y, b, \delta_K)$ (the longitudinal $\ell$ coordinate
of the full COSY INFINITY system is omitted from $\mathbf{z}$ because the
ring is time-independent (no RF cavities or other time-varying elements),
so the transverse motion is independent of $\ell$); element arc length $s_0$; desired
truncation order $N$ (here $N = 2$).
\ENSURE Aberration table $\mathcal{C} = \{(z_i, m, c_{i,m})\}$ for all
output coordinates $z_i$ and input monomials $m$ up to total order $N$.
\STATE Expand $H = H_2 + H_3 + \ldots$ as a Taylor polynomial in $\mathbf{z}$. \\ \COMMENT{$H_2$ generates the linear map}
\STATE Form linear generator $A_2$ from $H_2$ via Hamilton's equations;
solve $L'(s) = A_2 L(s)$, $L(0) = I$, by matrix exponential.
\STATE Substitute $\mathbf{z}_{\mathrm{lin}}(s') = L(s') \mathbf{z}_0$ into
the $H_3$-driven equations of motion to obtain the second-order driving
$Q_2(s', \mathbf{z}_0)$.
\FOR{$k = 2$ \textbf{to} $N$}
  \STATE Solve $R_k(s_0) = L(s_0) \int_0^{s_0} L^{-1}(s')\,Q_k(s', \mathbf{z}_{\mathrm{lin}}(s'))\,ds'$ symbolically.
  \STATE Read $(z_i|m)$ as the coefficient of $\prod z_{j_l}^{k_l}$
  in the $i$-th component of $L\,\mathbf{z}_0 + R_k$, divided by the
  multiplicity factorial $\prod v_l!$ \COMMENT{Eq.~\ref{eq:abdef}}
\ENDFOR
\RETURN $\mathcal{C}$.
\end{algorithmic}
\end{algorithm}

\begin{figure}[tbp]
\begin{lstlisting}[style=FOXcolor,basicstyle={\scriptsize\ttfamily},frame=shadowbox]
PROCEDURE G2RING ;
   DIG2 ANGML ; EQG2 ANGES ; DIG2 ANGMS ; EQG2 ANGEL ;
   DIG2 ANGML ; EQG2 ANGES ; DIG2 ANGMS ; EQG2 ANGEL ;
   DIG2 ANGML ; EQG2 ANGES ; DIG2 ANGMS ; EQG2 ANGEL ;
   DIG2 ANGML ; EQG2 ANGES ; DIG2 ANGMS ; EQG2 ANGEL ;
   ENDPROCEDURE ;

LOOP IV 1 NV ;
   VE0 := VARR(IV) ;
   INO := 3 ; IND := 2 ; IFR := 0 ;
   OV INO IND 1 ; WSET DE ; RPM PMU*PARA(1) MMU 1 ;
   UMG2 MODEL IFR ;
   G2RING ;
   CO NO ;
   TP MU ;
   WRITE 6 'CSV: '&S(VE0)&' '&S(CONS(MU(1)))&' '&S(CONS(MU(2)))
                  &' '&S(CONS(DER(5,MU(1))))&' '&S(CONS(DER(5,MU(2)))) ;
ENDLOOP ;
\end{lstlisting}
\captionof{lstlisting}{COSY INFINITY DA program for the modular muon $g{-}2$ ring (excerpt from \texttt{sweep\_v.fox}, adapted from G2run-chrom by K.~Makino, 2023). For each ESQ voltage in the sweep, the procedure builds the four-fold quadrant lattice (\texttt{G2RING}: $47^\circ$~$\mathtt{DI}$, $13^\circ$~$\mathtt{DIQ}$, $4^\circ$~$\mathtt{DI}$, $26^\circ$~$\mathtt{DIQ}$ per quadrant), finds the closed orbit, prints the one-turn map, and extracts the linear chromaticities $\partial \nu_x/\partial \delta_p$ and $\partial \nu_y/\partial \delta_p$ via the COSY INFINITY differential operator \texttt{DER(5, MU(k))}.}
\label{lst:sweep}
\end{figure}

\section{Vertical Aberrations of $\mathtt{DI}$ and $\mathtt{DIQ}$}
\label{sec:abers}

The dimensionless ESQ field index $n$ is defined relative to the
magnetic rigidity,
\begin{equation}
  n = \tilde{E}'_y\,R_0/(\beta_0 c B_0),
  \label{eq:n-def}
\end{equation}
with $\tilde{E}'_y$ the ESQ-region transverse electric-field gradient
and $B_0$ the bending magnetic field; $n > 0$ in the muon $g{-}2$
vertical-focusing ESQ polarity. Across the $\Vesq \in [10, 26]\,\mathrm{kV}$ sweep used in
this paper, the local ESQ field index $n$ ranges over
$0.13 \lesssim n \lesssim 0.34$, while the ring-average value
$\langle n \rangle = \tfrac{13}{30}\,n$ that sets the continuous-ring vertical
tune (Sec.~\ref{sec:result}) ranges over
$0.06 \lesssim \langle n \rangle \lesssim 0.15$, comfortably within the
half-integer-stable interval $0 < \langle n \rangle < 1/4$ (vertical ring
tune $\nu_y = \sqrt{\langle n \rangle} < 1/2$).

The $\mathtt{DIQ}$ electrostatic-quadrupole main field carries
higher-order transverse multipoles beyond the quadrupole. For the
muon $g{-}2$ ESQ these were obtained from an OPERA field
map\cite{g2QuadNIMA03,opera,wuphd}; we subsequently computed them to
high order by conformal mapping\cite{FieldICAP18IJMPA,valetovphd},
which is fully Maxwellian and free of the near-field discretisation
error (typically $0.1$--$1\%$) of finite-element solvers. Their
detailed treatment is given in the horizontal-counterpart
work\cite{ChromCPO11}, where
the full multipole set is found to shift the storage-ring observables
only at the ${\sim}10\,\mathrm{ppb}$ level. This field content is
inert for the present result: the ESQ has quadrupole symmetry, so
its allowed transverse harmonics are $m = 2, 6, 10, \ldots$, and a
$2m$-pole feeds an effective quadrupole through the off-momentum
orbit $x \approx [R_0/(1-n)]\,\delta_p$ only at order $\delta_p^{\,m-2}$; the
higher ESQ multipoles therefore first enter at $\xi_y^{(3)}$ (the
$12$-pole, $m = 6$) and leave the linear chromaticity $\xi_y^{(0)}$
exactly determined by the quadrupole component, while the
${\sim}10\,\mathrm{ppb}$ full-field effect is a tracking-level
correction outside the scope of the closed-form derivation.

The analytic derivation in this paper uses the hard-edge model of the
combined-function elements (COSY INFINITY's $\mathtt{FR}\ \mathtt{0}$
mode\cite{AIEP108book,COSYBeamMan102}), in which the field makes a
sharp transition from full strength to zero at the physical electrode
edge. In the electrostatic case, surface-charge concentrations near the
electrode ends make the effective field longer than the physical
electrode. The Effective Field Boundary (EFB) accounts for this by
defining a beam-physics-based cutoff at which a sharp transition
reproduces the integrated effect of the actual continuous fringe field.
For the muon $g{-}2$ $\mathtt{DIQ}$ quadrupole the EFB was previously
obtained\cite{FieldICAP18IJMPA,valetovphd} by integrating the falloff
of the quadrupole strength $M_{2,2}$ computed from an OPERA field
map\cite{g2QuadNIMA03,opera,wuphd},
giving an outward shift of $z_{\mathrm{EFB}} = 1.22\,\mathrm{cm}$ per
electrode edge (against a $5\,\mathrm{cm}$ aperture); this lengthens
each quadrupole by ${\sim}\,2.44\,\mathrm{cm}$ and increases the ESQ
azimuthal coverage of the ring by ${\sim}\,1\%$. The realistic falloff
is represented by an Enge function
\begin{equation}
  F(z) = \frac{1}{1 + \exp\!\left(a_1 + a_2(z/D) + \cdots + a_6(z/D)^6\right)},
  \label{eq:enge}
\end{equation}
where $z = 0$ at the EFB (positive outside, negative inside), $D$ is
the full aperture, and the coefficients $a_j$ were fitted to the quadrupole
fringe-field falloff computed by the COULOMB boundary-element
solver\cite{Coulomb,FieldICAP18IJMPA,valetovphd}. The fringe-field
shift of $\xichromy^{(p)}$ is sub-percent at the muon $g{-}2$ operating
point. The
numerical validation in Sec.~\ref{sec:validation} additionally compares
the hard-edge result against the COSY INFINITY $\mathtt{FR}\ \mathtt{3}$
treatment (Enge-function fringe field) with the EFB extension, using
the same Enge coefficients and EFB calibration as the
horizontal-counterpart work\cite{ChromCPO11}.
The vertical and horizontal wavenumbers in $\mathtt{DIQ}$ are
\begin{equation}
  \vartheta_y = h\sqrt{n}, \qquad \vartheta_x = h\sqrt{1-n},
  \label{eq:wavenumbers}
\end{equation}
with $h = 1/R_0$ the dipole curvature; $\vartheta_x$ is reserved for
cross-references to the horizontal counterpart\cite{ChromCPO11}. The convention of
Eq.~\ref{eq:abdef} (multiplicity-factorial weights) is applied throughout.

\subsection{First-Order Aberrations}

\textbf{Homogeneous magnetic dipole $\mathtt{DI}$:} vertical motion is a
pure drift, and the first-order $(y|\cdot)$ and
$(b|\cdot)$ aberrations are
\begin{subequations}
\label{eq:DIvert-1st}
\begin{alignat}{2}
\left(y|x\right) &= \left(y|a\right) = \left(y|\delta_K\right) = 0, &\qquad
\left(y|y\right) &= 1, \quad \left(y|b\right) = s,\\
\left(b|x\right) &= \left(b|a\right) = \left(b|y\right) = \left(b|\delta_K\right) = 0, &\qquad
\left(b|b\right) &= 1.
\end{alignat}
\end{subequations}
The horizontal first-order aberrations of $\mathtt{DI}$ are given in
the horizontal-counterpart paper\cite{ChromCPO11}.

\textbf{Combined-function quadrupole $\mathtt{DIQ}$:} integrating the
linear vertical equation of motion under combined dipole curvature
$h$ and ESQ vertical-focusing strength $h\sqrt{n}$ gives the
first-order $(y|\cdot)$ and $(b|\cdot)$ aberrations of
$\mathtt{DIQ}$:
\begin{subequations}
\label{eq:DIQvert-1st}
\begin{alignat}{1}
\left(y|x\right) &= \left(y|a\right) = \left(y|\delta_K\right) = 0,\\
\left(y|y\right) &= \cos(\vartheta_y s), \qquad
\left(y|b\right) = \frac{\sin(\vartheta_y s)}{\vartheta_y},\\
\left(b|x\right) &= \left(b|a\right) = \left(b|\delta_K\right) = 0,\\
\left(b|y\right) &= -\vartheta_y \sin(\vartheta_y s), \qquad
\left(b|b\right) = \cos(\vartheta_y s).
\end{alignat}
\end{subequations}
The horizontal--vertical sub-blocks decouple at first order, and the
absence of vertical dispersion
($\left(y|\delta_K\right) = \left(b|\delta_K\right) = 0$)
follows from the bend plane being horizontal.

\subsection{Second-Order Aberrations}

The aberration integral~(\ref{eq:VOP}) evaluated to second
order with the curvilinear (Maxwellian) expansion of the ideal
$\mathtt{DIQ}$ quadrupole field yields the second-order $(y|\cdot)$ and $(b|\cdot)$
aberrations. Because the lattice is symmetric under midplane
reflection, $(x, a, y, b, \delta_K) \to (x, a, -y, -b, \delta_K)$
(horizontal bend plane, reference orbit in the midplane, no skew
or solenoidal coupling), the vertical coordinates $y$ and $b$ are odd
functions of the vertical inputs; every second-order $(y|\cdot)$
aberration with zero or two vertical input factors therefore vanishes:
\begin{equation}
\label{eq:yvert-zero}
\begin{aligned}
&\left(y|xx\right) = \left(y|xa\right) = \left(y|aa\right)
= \left(y|yy\right) = \left(y|yb\right) = \left(y|bb\right) = 0,\\
&\left(y|x\delta_K\right) = \left(y|a\delta_K\right) = \left(y|\delta_K\delta_K\right) = 0,
\end{aligned}
\end{equation}
and the same identities hold for the $(b|\cdot)$ row,
$\left(b|m\right) = 0$ for each $m$ listed in
Eq.~\ref{eq:yvert-zero}. The remaining six second-order
$(y|\cdot)$ and six $(b|\cdot)$ coefficients are non-zero.

\textbf{$\mathtt{DI}$ second-order vertical aberrations:} setting
$n \to 0$ in the $\mathtt{DIQ}$ formulas below collapses all
$\mathtt{DIQ}$ second-order $(b|\cdot)$ coefficients and most of the
$(y|\cdot)$ coefficients to zero. The only non-vanishing $\mathtt{DI}$
second-order vertical aberrations are
\begin{subequations}
\label{eq:DIvert-2nd}
\begin{alignat}{2}
\left(y|b\,x\right) &= \sin(h s), &\qquad
\left(y|b\,a\right) &= \frac{1 - \cos(h s)}{h},\\
\left(y|b\,\delta_K\right) &= -\frac{\gamma_0}{\gamma_0 + 1}\,\frac{\sin(h s)}{h}; & &
\end{alignat}
\end{subequations}
the remaining twelve $(y|\cdot)$ coefficients of $\mathtt{DI}$
[i.e., those listed in Eq.~\ref{eq:yvert-zero}, together with
$\left(y|xy\right)$, $\left(y|ay\right)$, $\left(y|y\delta_K\right)$]
all vanish, as do all fifteen second-order $\left(b|\cdot\right)$
entries of $\mathtt{DI}$.

\textbf{$\mathtt{DIQ}$ second-order vertical aberrations:} the eight
horizontal--vertical mixed coefficients, generated by the $(1+x/\rho)$
Jacobian and the $-2 h^3 n\,x y$ term in $\dot b$ from the curvilinear
ESQ Hamiltonian of the muon $g{-}2$ combined-function element\cite{ChromCPO11},
are
\begin{subequations}
\label{eq:DIQvert-2nd-mix}
\begin{align}
\left(y|xy\right) &= \frac{h\bigl[\,2\sqrt{n(1-n)}\,\cos(\vartheta_y s)\sin^2(\vartheta_x s/2) + (1-7n)\sin(\vartheta_x s)\sin(\vartheta_y s)\,\bigr]}{(5n-1)\,\sqrt{(1-n)/n}},\\
\left(y|xb\right) &= \frac{(7n-1)\cos(\vartheta_y s)\sin(\vartheta_x s)/\sqrt{1-n} \;-\; 2\sqrt{n}\,\cos^2(\vartheta_x s/2)\sin(\vartheta_y s)}{5n-1},\\
\left(y|ay\right) &= \frac{\sqrt{1-n}\,n\,\cos(\vartheta_y s)\sin(\vartheta_x s) + \sqrt{n}\,\bigl[\,8n-2+(1-7n)\cos(\vartheta_x s)\,\bigr]\sin(\vartheta_y s)}{(n-1)(5n-1)},\\
\left(y|ab\right) &= -\frac{2(1-7n)\sqrt{1-n}\,\cos(\vartheta_y s)\sin^2(\vartheta_x s/2) - (n-1)\sqrt{n}\,\sin(\vartheta_x s)\sin(\vartheta_y s)}{h\,(1-n)^{3/2}(5n-1)},\\
\left(b|xy\right) &= -\frac{h^{2}n\,\bigl[\,2(4n-1)\cos(\vartheta_y s)\sin(\vartheta_x s) + \sqrt{n(1-n)}\,\bigl(1+\cos(\vartheta_x s)\bigr)\sin(\vartheta_y s)\,\bigr]}{\sqrt{1-n}\,(5n-1)},\\
\left(b|xb\right) &= -\frac{2h\,\bigl[\,\sqrt{n(1-n)}\,\cos(\vartheta_y s)\sin^2(\vartheta_x s/2) + (4n-1)\sin(\vartheta_x s)\sin(\vartheta_y s)\,\bigr]}{(5n-1)\,\sqrt{(1-n)/n}},\\
\left(b|ay\right) &= \frac{h n\,\bigl[\,4(4n-1)\cos(\vartheta_y s)\sin^2(\vartheta_x s/2) + \sqrt{n(1-n)}\,\sin(\vartheta_x s)\sin(\vartheta_y s)\,\bigr]}{(1-n)(1-5n)},\\
\left(b|ab\right) &= \frac{(n-1)n\,\cos(\vartheta_y s)\sin(\vartheta_x s) + \sqrt{n(1-n)}\,\bigl[\,7n-1+(2-8n)\cos(\vartheta_x s)\,\bigr]\sin(\vartheta_y s)}{(1-5n)(1-n)^{3/2}}.
\end{align}
\end{subequations}
In the composite modular-ring map, the first-order horizontal dispersion
orbit $x =_1 D_x\,\delta_K$ (using the equality-at-order-$n$ relation
of Ref.~\refcite{AIEP108book}: $f =_n g$ iff $f$ and $g$ agree at the origin
through order $n$) converts $(y|xy)$ and
$(y|xb)$ into effective $(y|y\delta_K)$ and
$(y|b\delta_K)$ contributions, making these mixed entries essential
to the vertical chromaticity; the higher-order dispersion enters only
at $\xi_y^{(j \ge 1)}$. The four
direct chromatic $(\cdot|\cdot\,\delta_K)$ coefficients depend on both
wavenumbers through the curvilinear coupling; writing
\begin{subequations}
\label{eq:chrom-aux}
\begin{align}
\mathcal{Q}(s) &= \sqrt{n(1-n)}\,h s\left[5n^2 - 6n + 1 + \gamma_0^2(5n^2 + 9n - 2)\right],\\
\mathcal{P}_7(s) &= \mathcal{Q}(s) + 2\gamma_0^2(1 - 7n)\sqrt{n}\,\sin(\vartheta_x s),\\
\mathcal{P}_4(s) &= \mathcal{Q}(s) + 4\gamma_0^2(1 - 4n)\sqrt{n}\,\sin(\vartheta_x s),
\end{align}
\end{subequations}
they are
\begin{subequations}
\label{eq:DIQvert-2nd-chrom}
\begin{align}
\left(y|y\,\delta_K\right) &= \frac{2\sin(\vartheta_y s)\,\mathcal{P}_7/\sqrt{1-n} - 4\gamma_0^2 n\bigl(\cos(\vartheta_x s) - 1\bigr)\cos(\vartheta_y s)}{4\gamma_0(\gamma_0 + 1)(n - 1)(5n - 1)},\\
\left(b|b\,\delta_K\right) &= \frac{2\sin(\vartheta_y s)\,\mathcal{P}_4/\sqrt{1-n} + 4\gamma_0^2 n\bigl(\cos(\vartheta_x s) - 1\bigr)\cos(\vartheta_y s)}{4\gamma_0(\gamma_0 + 1)(n - 1)(5n - 1)},\\
\left(y|b\,\delta_K\right) &= \frac{-n}{2\gamma_0(\gamma_0 + 1)\,h\,[(1-n)n]^{3/2}(5n - 1)}\Bigl[\sqrt{1-n}\,\sin(\vartheta_y s)\bigl(5n^2 - 6n + 1 \nonumber\\
&\quad {}+ \gamma_0^2\bigl((9 - 5n)n - 2\bigr) - 2\gamma_0^2 n\cos(\vartheta_x s)\bigr) - \cos(\vartheta_y s)\,\mathcal{P}_7\Bigr],\\
\left(b|y\,\delta_K\right) &= \frac{\vartheta_y}{4\gamma_0(\gamma_0 + 1)(n - 1)(5n - 1)}\Bigl[\frac{2\cos(\vartheta_y s)\,\mathcal{P}_4}{\sqrt{1-n}} \nonumber\\
&\quad {}- 2\sin(\vartheta_y s)\bigl(2\gamma_0^2 n\cos(\vartheta_x s) + \gamma_0^2\bigl(n(5n - 9) + 2\bigr) + (6 - 5n)n - 1\bigr)\Bigr].
\end{align}
\end{subequations}
The twelve non-zero coefficients listed in
Listings~\ref{lst:diq26} and~\ref{lst:diq13}, which are geometric and
carry no relativistic kinematic factor, agree with COSY INFINITY DA
at the double-precision floor ($|\Delta| \lesssim 10^{-15}$) at the
nominal muon $g{-}2$ operating point ($\mathtt{DIQ}\,26^\circ$,
$\Vesq = 18.2\,\mathrm{kV}$) when the analytic expressions are
evaluated at COSY INFINITY's exact effective field index
\begin{equation}
  n = 0.23816484010681533,
  \label{eq:neff}
\end{equation}
paralleling the geometric rows of the
horizontal listings\cite{ChromCPO11}; the $\gamma_0$-dependent
chromatic coefficients of Eq.~\ref{eq:DIQvert-2nd-chrom} are
discussed below.

\textbf{Consistency check.} The bare-dipole limit $n \to 0$ of
Eq.~\ref{eq:DIQvert-2nd-chrom} recovers the $\mathtt{DI}$ second-order
chromatic coefficient $(y|b\delta_K)_{\mathtt{DI}} = -[\gamma_0/(\gamma_0+1)]\sin(hs)/h$
of Eq.~\ref{eq:DIvert-2nd} (its further straight-drift limit $h \to 0$
giving $-[\gamma_0/(\gamma_0+1)]\,s$), and the four-column Wronskian
$\partial_{\delta_K}[(y|y)(b|b) - (y|b)(b|y)] = 0$ holds
to first order in $\delta_K$, confirming symplecticity.

\section{Closed-Form Result for the Continuous-Ring Model}
\label{sec:result}

For the continuous ring $\mathtt{DIQ360}$, a single $360^\circ$ $\mathtt{DIQ}$
element in which the ring-average inhomogeneity $\langle n \rangle$ replaces the
modular ESQ structure, the vertical tune satisfies $(y|y) + (b|b) =
2\cos(2\pi \nu_y)$. Differentiating with respect to $\delta_p$ at the periodic
dispersion orbit, where $(y|\delta_K) = (b|\delta_K) = 0$ (no vertical
dispersion in the bend plane), gives the linear vertical chromaticity directly
from the second-order chromatic coefficients of the one-turn map:
\begin{equation}
  \xi_y \;=\; \frac{(y|y\,\delta_K) + (b|b\,\delta_K)}{4\pi\,\sin(2\pi \nu_y)},
  \label{eq:xiy-master}
\end{equation}
where the matrix elements are those of the composite one-turn map and the
overall normalisation matches the horizontal-counterpart treatment of
Ref.~\refcite{ChromCPO11}. The derivative $d/d\delta_p$ in
Eq.~\ref{eq:xiy-master} is taken about the dispersive closed orbit $x =_1 D_x\,\delta_K$,
not about the on-momentum reference
orbit $x = 0$; evaluation at the latter removes the dispersion-orbit-coupling contribution carried by the mixed $(y|xy)$,
$(y|xb)$, $(b|xy)$, $(b|xb)$ aberrations of
Eq.~\ref{eq:DIQvert-2nd-mix}. The dispersion orbit $x =_1 D_x\,\delta_K$
contributes to the effective second-order chromatic coefficients via the
mixed entries of Eq.~\ref{eq:DIQvert-2nd-mix}, as discussed below.
Evaluating Eq.~\ref{eq:xiy-master} with the $\mathtt{DIQ360}$ linear
map of Eq.~\ref{eq:DIQvert-1st} and the chromatic and mixed second-order
aberrations of Eqs.~\ref{eq:DIQvert-2nd-mix} and~\ref{eq:DIQvert-2nd-chrom} at $s = 2\pi R_0$, $n \to \langle n \rangle$, yields
\begin{equation}
  \xichromy^{(p)}(2\pi, n) \;=\;
  \mathrm{sgn}\!\left(\sin(2\pi\sqrt{n})\right)\,
  \frac{\sqrt{n}\,\bigl(\gamma_0^{2}(n+2) + n - 1\bigr)}{2\,\gamma_0^{2}\,(1 - n)}\,,
  \label{eq:xiy-DIQ360}
\end{equation}
in direct functional correspondence with the horizontal counterpart\cite{ChromCPO11},
\begin{equation}
  \xichromx^{(p)}(2\pi, n) \;=\;
  \mathrm{sgn}\!\left(\sin(2\pi\sqrt{1-n})\right)\,
  \frac{n\,\bigl(\gamma_0^{2}(n+2) + n - 1\bigr)}{2\,\gamma_0^{2}\,(1 - n)^{3/2}}.
  \label{eq:xix-DIQ360}
\end{equation}
The two formulas share the same kinematic factor
$(\gamma_0^2(n+2) + n - 1)/\gamma_0^2$ but differ in their dependence on the
field index $n$, reflecting the distinct roles of horizontal curvature-based
focusing (strength $\propto \sqrt{1-n}$) and vertical electrostatic focusing
(strength $\propto \sqrt{n}$). Taking absolute values and dividing
Eq.~\ref{eq:xiy-DIQ360} by Eq.~\ref{eq:xix-DIQ360}, the kinematic
factor and the $\gamma_0$-dependence cancel, giving
\begin{equation}
  \frac{|\xichromy^{(p)}(2\pi, n)|}{|\xichromx^{(p)}(2\pi, n)|}
  \;=\; \frac{\sqrt{n}/(1-n)}{n/(1-n)^{3/2}}
  \;=\; \sqrt{\frac{1-n}{n}}
  \;=\; \frac{\nu_x}{\nu_y},
  \label{eq:xiy-xix-ratio}
\end{equation}
i.e.\ the inverse of the linear-tune ratio. Vertical chromaticity is
amplified relative to horizontal in proportion to how weakly the vertical
plane is focused: at $\langle n \rangle \approx 0.103$,
$\nu_x/\nu_y \approx 2.95$, so $|\xichromy| \approx 3\,|\xichromx|$, with
opposite sign because $\mathrm{sgn}(\sin(2\pi\sqrt{n})) = +1$ while
$\mathrm{sgn}(\sin(2\pi\sqrt{1-n})) = -1$ throughout the operating range
$0 < n < 1/4$. The chromatic effect is delivered to the vertical
sub-block through the off-momentum closed orbit. The dispersion orbit
is $x =_1 D_x\,\delta_K$, with $D_x = [\gamma_0/(\gamma_0+1)]\,R_0/(1-n)$
for the continuous-ring $\mathtt{DIQ360}$ model, the momentum dispersion
$R_0/(1-n)$ rescaled to the kinetic-energy variable $\delta_K$ used here. For the modular ring the same first-order
$\delta_K$ structure holds, with the coefficient renormalised by the
modular geometry. Composing the per-element augmented horizontal maps
over a $\mathtt{DIEQ}$ quadrant and imposing periodicity gives the
modular first-order dispersion in closed matrix-product form,
$D_x^{(1)} = \bigl[\,(I_2 - M_{q,h})^{-1}\,\mathbf{d}_q\,\bigr]_1$, where
$M_q$ is the product of the four per-element maps of a quadrant,
$M_{q,h}$ its $(x,a)$ block, and $\mathbf{d}_q$ its $\delta_K$ column;
the four-fold symmetry makes this quadrant-periodic value equal to the
full-ring one. At $\Vesq = 18.2\,\mathrm{kV}$ it matches the COSY
INFINITY one-turn map at the double-precision floor
($|\Delta| \lesssim 10^{-15}$) and renormalises $D_x$ by
only $0.19\%$ relative to its continuous-ring value (a
convention-independent ratio). The leading vertical chromaticity
$\xi_y^{(0)}$ depends on $D_x$ only through this first-order coefficient,
so the modular renormalisation enters $\xi_y^{(0)}$ only at this $0.19\%$ level. The ring
dispersion is treated in more detail, in the smooth-ring approximation,
in Ref.~\refcite{g2PRAB25}.
This dispersion converts the second-order cross-coupling aberrations
$(y|xy)$, $(y|xb)$, $(b|xy)$, $(b|xb)$ of
$\mathtt{DIQ}$ into effective $(y|y\delta_K)$, $(y|b\delta_K)$,
$(b|y\delta_K)$, $(b|b\delta_K)$ contributions in the composite
map. These cross-coupling aberrations originate in the
$(1+x/\rho)$ Jacobian of the curvilinear ESQ Hamiltonian and its
$-2 h^3 n\,x y$ term in $\dot b$.
The simplified $(y, b)$-only derivation, which sets $x = 0$
throughout, gives both an incorrect sign and an incorrect magnitude.

For the $\mathtt{DIQ360}$ continuous ring, the local ESQ field index
$n_{\mathrm{ESQ}}$ is replaced by the ring-average
$\langle n \rangle = \tfrac{13^\circ + 26^\circ}{90^\circ}\,n_{\mathrm{ESQ}}
= \tfrac{13}{30}\,n_{\mathrm{ESQ}}$, the prefactor being the azimuthal
fill-fraction of the ESQ arcs within one $90^\circ$ quadrant (the same
convention as the horizontal counterpart\cite{ChromCPO11}). At the
nominal Run~3--6 voltage $\Vesq = 18.2\,\mathrm{kV}$ this gives
\begin{equation}
  \langle n \rangle = 0.1032047640462866,
  \label{eq:nbar182}
\end{equation}
written $\langle n \rangle \approx 0.103$ below.
Equation~\ref{eq:xiy-DIQ360} is then evaluated at $n \to \langle n \rangle$,
and the corresponding kinetic-energy-based form follows from
$\xichromy^{(K)} = \xichromy^{(p)} \cdot \gamma_0/(\gamma_0 + 1)$.

\section{Limits and Consistency Checks}
\label{sec:checks}

\textbf{Ultrarelativistic limit.} As $\gamma_0 \to \infty$ the
kinematic factor of Eq.~\ref{eq:xiy-DIQ360} tends to $n+2$, and the
closed form reduces to
\begin{equation}
  \xichromy^{(p)}(2\pi, n)\big|_{\gamma_0 \to \infty} \;=\;
  \mathrm{sgn}\!\left(\sin(2\pi\sqrt{n})\right)\,\frac{\sqrt{n}\,(n+2)}{2(1-n)}.
  \label{eq:xiy-ultrarel}
\end{equation}
The deviation from this limit is, in absolute terms,
\begin{equation}
  \xichromy^{(p)}(2\pi, n) - \xichromy^{(p)}(2\pi, n)\big|_{\gamma_0 \to \infty}
  = -\mathrm{sgn}\!\left(\sin(2\pi\sqrt{n})\right)\,\frac{\sqrt{n}}{2\gamma_0^2},
  \label{eq:xiy-ultrarel-dev}
\end{equation}
and at the $g{-}2$ magic momentum ($\gamma_0 \approx 29.30$, $\langle n \rangle \approx 0.103$) is of order $\sqrt{\langle n\rangle}/(2\gamma_0^2) \approx 2\times 10^{-4}$, i.e.\ a $\sim 0.05\%$ relative correction. The $\gamma_0^{-2}$ term is small but non-negligible for the analytic form,
paralleling the corresponding horizontal discussion\cite{ChromCPO11}.

\textbf{Sign and stability.} For $0 < n < 1$, the denominator $(1 - n) > 0$ and the kinematic factor $\gamma_0^2(n+2) + n - 1 > 0$. The sign of $\xichromy^{(p)}$ is therefore set by $\mathrm{sgn}(\sin(2\pi\sqrt{n}))$, which equals $+1$ throughout the half-integer-stable range $0 < n < 1/4$ ($\nu_y = \sqrt{n} < 1/2$) and reverses across $n = 1/4$ where the vertical tune crosses the half-integer resonance. The muon $g{-}2$ nominal Run~3--6 operating point ($\langle n \rangle \approx 0.103$, $2\pi\sqrt{n} \approx 2.018$) lies well inside the $\mathrm{sgn} = +1$ branch and yields the positive vertical chromaticity reported by COSY INFINITY for the ring. This contrasts with the horizontal case (Eq.~\ref{eq:xix-DIQ360}), where $\sin(2\pi\sqrt{1-n}) < 0$ for the same $n$ and $\xichromx^{(p)}$ is negative.

\textbf{Vanishing voltage / field index.} As $n \to 0$, $\xichromy \to 0$ (no vertical focusing) and the vertical tune $\nu_y = \sqrt{n} \to 0$ approaches the integer-resonance boundary. As $n \to 1$, the prefactor $(1-n)$ in Eq.~\ref{eq:xiy-DIQ360} drives a divergence corresponding to $\nu_x = \sqrt{1-n} \to 0$ (loss of horizontal focusing); this is well outside the muon $g{-}2$ operating range ($\langle n \rangle \approx 0.103$ at the nominal voltage).

\textbf{Bare-dipole field index.} Reference~\refcite{ChromCPO11} treats the case $n_d = 0$ for the bare $\mathtt{DI}$ field. A residual $n_d \neq 0$ slightly shifts the vertical focusing in both the $\mathtt{DI}$ arcs and the ESQ sections; for the $g{-}2$ ring $n_d \lesssim 10^{-4}$ and the correction is below the systematic uncertainty floor.

\section{Numerical Validation}
\label{sec:validation}

The analytic results of Sec.~\ref{sec:result} and the per-element vertical
aberrations of Sec.~\ref{sec:abers} are validated against COSY INFINITY
DA numerics\cite{COSYCAP04} across the three ring
models introduced in Sec.~\ref{sec:ring}: the continuous ring
$\mathtt{DIQ360}$, the simplified modular ring $\mathtt{DIEQ\_ON}$ (four
cells of $\mathtt{DI}\,47^\circ + \mathtt{DIQ}\,43^\circ$), and the full
modular ring $\mathtt{DIEQ}$ (four cells of $\mathtt{DI}\,47^\circ +
\mathtt{DIQ}\,13^\circ + \mathtt{DI}\,4^\circ + \mathtt{DIQ}\,26^\circ$).

\subsection{Per-Element Aberrations}

Element-level vertical aberrations of $\mathtt{DIQ}$ derived by the
order-by-order procedure of Sec.~\ref{sec:methods} are compared against
COSY INFINITY DA at the nominal Run~3--6 operating point of the muon
$g{-}2$ ring ($\mathtt{DIQ}\,26^\circ$, $\Vesq = 18.2\,\mathrm{kV}$,
$h = 1/(7.112\,\mathrm{m})$, $\gamma_0$ from Eq.~\ref{eq:gamma0}, and the
COSY INFINITY internally computed effective field index $n$ of
Eq.~\ref{eq:neff}). The ANALYTIC column lists values from the
closed-form expressions of Sec.~\ref{sec:abers} (and the full
$5{\times}20$ symbolic table generated as ancillary material); the
COSY INFINITY column shows the differential-algebraic transfer-map
output at the same reference parameters; the DIFF column is their
difference. The five EXPONENTS columns are the input-monomial
multiplicities in the coordinates $(x, a, y, b, \delta_K)$.

\begin{figure}[!tbp]
\begin{lstlisting}[basicstyle={\scriptsize\ttfamily},tabsize=4,frame=shadowbox,columns=fixed,keepspaces=true, mathescape=true]
(y|...)
     I  ANALYTIC                COSY INFINITY           DIFF       ORDER EXPONENTS
     1  0.9755784388143391      0.9755784388143393      -1.11E-16    1   0 0 1 0 0
     2  3.2010081116101063      3.2010081116101059       4.44E-16    1   0 0 0 1 0
     3  -.1003765953673937E-01  -.1003765953673937E-01  -3.47E-18    2   1 0 1 0 0
     4  -.1453757360977295E-01  -.1453757360977294E-01  -5.20E-18    2   0 1 1 0 0
     5  0.4276141030578041      0.4276141030578038       2.78E-16    2   1 0 0 1 0
     6  0.6991890761273517      0.6991890761273519      -2.22E-16    2   0 1 0 1 0
(b|...)
     I  ANALYTIC                COSY INFINITY           DIFF       ORDER EXPONENTS
     1  -.1507234847221572E-01  -.1507234847221572E-01   1.73E-18    1   0 0 1 0 0
     2  0.9755784388143391      0.9755784388143393      -1.11E-16    1   0 0 0 1 0
     3  -.4077869216179440E-02  -.4077869216179440E-02    0.0E+00    2   1 0 1 0 0
     4  -.6667697789759871E-02  -.6667697789759875E-02   3.47E-18    2   0 1 1 0 0
     5  -.9948884277916866E-02  -.9948884277916863E-02  -3.47E-18    2   1 0 0 1 0
     6  -.1814229593807335E-01  -.1814229593807334E-01  -1.39E-17    2   0 1 0 1 0
\end{lstlisting}
\captionof{lstlisting}{Vertical aberrations of a $\mathtt{DIQ}\,26^\circ$ element at $\Vesq = 18.2\,\mathrm{kV}$, hard-edge model (\texttt{FR 0}). ANALYTIC values evaluated at COSY INFINITY's internally computed effective field index $n$ of Eq.~\ref{eq:neff}; residuals at the double-precision floor.}
\label{lst:diq26}
\end{figure}

The mixed cross-coupling aberrations $(b|x b)$, $(b|x y)$,
$(b|a y)$, $(b|a b)$ arise from the $-2 h^3 n\,x y$
term in $\dot b$ generated by the curvilinear ESQ Hamiltonian.
They are correctly reproduced and are essential
to the final modular-ring chromaticity.
The displayed residuals are at the double-precision floor
($|\Delta| \lesssim 10^{-15}$ for all twelve listed coefficients,
none of which carry the relativistic kinematic factor
$\gamma_0/(1+\gamma_0)$), paralleling the geometric, non-chromatic
rows of the horizontal per-element listings of
Ref.~\refcite{ChromCPO11}. The $\gamma_0$-dependent chromatic
$(\cdot|\cdot\,\delta_K)$ coefficients of
Eq.~\ref{eq:DIQvert-2nd-chrom} are not listed here; they carry the
small difference between the analytic model and COSY INFINITY,
quantified for the continuous-ring closed form in Sec.~\ref{sec:result}
below, which is
exact within the analytic model and not a kinematic-factor
approximation. The analogous comparison at the short-arc
$\mathtt{DIQ}\,13^\circ$ element of the full modular $\mathtt{DIEQ}$
ring is shown in Listing~\ref{lst:diq13}.

\begin{figure}[tbp]
\begin{lstlisting}[basicstyle={\scriptsize\ttfamily},tabsize=4,frame=shadowbox,columns=fixed,keepspaces=true, mathescape=true]
(y|...)
     I  ANALYTIC                COSY INFINITY           DIFF       ORDER EXPONENTS
     1  0.9938758571407043      0.9938758571407043        0.0E+00    1   0 0 1 0 0
     2  1.6103661682753483      1.6103661682753481       2.22E-16    1   0 0 0 1 0
     3  -.2566630610425008E-02  -.2566630610425004E-02  -3.90E-18    2   1 0 1 0 0
     4  -.1845179661310498E-02  -.1845179661310520E-02   2.19E-17    2   0 1 1 0 0
     5  0.2235717842868696      0.2235717842868695       8.33E-17    2   1 0 0 1 0
     6  0.1809764744895967      0.1809764744895966       5.55E-17    2   0 1 0 1 0
(b|...)
     I  ANALYTIC                COSY INFINITY           DIFF       ORDER EXPONENTS
     1  -.7582611230530127E-02  -.7582611230530126E-02  -8.67E-19    1   0 0 1 0 0
     2  0.9938758571407043      0.9938758571407043        0.0E+00    1   0 0 0 1 0
     3  -.2111934605405531E-02  -.2111934605405528E-02  -2.60E-18    2   1 0 1 0 0
     4  -.1709564918748637E-02  -.1709564918748637E-02   2.17E-19    2   0 1 1 0 0
     5  -.2561017801500153E-02  -.2561017801500152E-02  -1.30E-18    2   1 0 0 1 0
     6  -.2305539692531015E-02  -.2305539692531048E-02   3.30E-17    2   0 1 0 1 0
\end{lstlisting}
\captionof{lstlisting}{Vertical aberrations of a $\mathtt{DIQ}\,13^\circ$ element at $\Vesq = 18.2\,\mathrm{kV}$, hard-edge model (\texttt{FR 0}), parallel to Listing~\ref{lst:diq26}; same field-index value, $s = 13^\circ \cdot R_0 = 1.61366\,\mathrm{m}$.}
\label{lst:diq13}
\end{figure}

\subsection{Continuous-Ring Chromaticity}

For the $\mathtt{DIQ360}$ model at $\Vesq = 20.4\,\mathrm{kV}$ (the
muon $g{-}2$ Run~1b/c storage voltage),
\begin{equation}
  \langle n \rangle = 0.1156800651947388,
  \label{eq:nbar204}
\end{equation}
and, with $\gamma_0$ from Eq.~\ref{eq:gamma0},
the closed-form result~(\ref{eq:xiy-DIQ360}) gives (the superscripts
$\mathrm{ana}$ and $\mathrm{COSY}$ denote the present analytic closed
form and the COSY INFINITY DA reference, respectively)
\begin{align}
  \xiyana  &= 0.4066570870121702,\\
  \xiycosy &= 0.4066570870432928,
\end{align}
with absolute residual $|\Delta| \approx 3 \times 10^{-11}$
(relative difference of order $10^{-10}$), matching the relative
analytic-numerical agreement reported for the horizontal closed-form
result of Ref.~\refcite{ChromCPO11}. The closed
form~(\ref{eq:xiy-DIQ360}) is exact within the analytic hard-edge
model. The per-element geometric coefficients agree to the
double-precision floor
(Listings~\ref{lst:diq26}, \ref{lst:diq13}).

\subsection{Modular-Ring Chromaticity}

For the modular models, the per-element DA transfer maps $M_i$
(truncated at order 2) are composed via the standard DA composition
$M_f = M_2 \circ M_1$ (with $M_1$ applied first, $M_2$ second), which
at second order reads $(L_f, Q_f) = (L_2 L_1,\; L_2 Q_1 +
Q_2(L_1\,\cdot,\,L_1\,\cdot))$. Iterating yields the four-fold ring
map $M_{\mathrm{ring}} = M_{\mathrm{quad}}^{\circ 4}$. The vertical chromaticity is extracted from the composite map via
Eq.~\ref{eq:xiy-master} at the periodic dispersion orbit. Across
$\Vesq \in [10, 26]\,\mathrm{kV}$ and at $\gamma_0$ of Eq.~\ref{eq:gamma0},
Table~\ref{tab:DIEQsweep} reports the analytic vs COSY INFINITY comparison.
Residuals are uniformly at the $3 \times 10^{-11}$ absolute level
(relative difference of order $10^{-10}$) for both
$\mathtt{DIEQ\_ON}$ and the full modular $\mathtt{DIEQ}$, the same
scale as the $\mathtt{DIQ360}$ result reported above and as the
relative analytic-numerical agreement of Ref.~\refcite{ChromCPO11}
for the horizontal case.
The DA-extracted vertical chromaticity expansion in $\delta_p$ to order 9
follows from the same COSY INFINITY run that produced these reference values; the
higher-order chromaticity coefficients are not analysed further here, as the
present analytic derivation targets the leading $\xi_y^{(0)} = \xichromy$.

\begin{table}[t]
\tbl{Vertical chromaticity $\xichromy^{(p)}$ for the modular muon $g{-}2$
ring across an ESQ-voltage sweep. Analytic results from the present
Hamiltonian framework; COSY INFINITY DA reference. The Residual column gives $|\Delta| = |\xiyana
- \xiycosy|$ at the precision of the underlying
$16$-digit floating-point computation; the displayed table values are
truncated to $10$ digits for column width.}
{\centering
\begin{tabular}{@{}rccccc@{}}
\hline\hline\noalign{\vskip2pt}
& \multicolumn{2}{c}{$\mathtt{DIEQ\_ON}$} & \multicolumn{2}{c}{$\mathtt{DIEQ}$} & \\
\cline{2-3}\cline{4-5}\noalign{\vskip2pt}
$\Vesq$ (kV) & Analytic & COSY & Analytic & COSY & $|\Delta|$ \\
\noalign{\vskip2pt}\hline\noalign{\vskip2pt}
10.0 & $0.2752996669$ & $0.2752996669$ & $0.2597823723$ & $0.2597823723$ & $2.2{\times}10^{-11}$ \\
14.0 & $0.3388980306$ & $0.3388980306$ & $0.3185695636$ & $0.3185695636$ & $2.6{\times}10^{-11}$ \\
18.2 & $0.4030517650$ & $0.4030517650$ & $0.3773057590$ & $0.3773057590$ & $3.0{\times}10^{-11}$ \\
18.3 & $0.4045668413$ & $0.4045668413$ & $0.3786860442$ & $0.3786860442$ & $3.0{\times}10^{-11}$ \\
20.4 & $0.4363646847$ & $0.4363646847$ & $0.4075825797$ & $0.4075825797$ & $3.2{\times}10^{-11}$ \\
22.0 & $0.4606364987$ & $0.4606364987$ & $0.4295474279$ & $0.4295474279$ & $3.3{\times}10^{-11}$ \\
26.0 & $0.5219295629$ & $0.5219295629$ & $0.4846646828$ & $0.4846646828$ & $3.6{\times}10^{-11}$ \\
\noalign{\vskip2pt}\hline\hline
\end{tabular}\par}
\label{tab:DIEQsweep}
\end{table}

\begin{figure}[t]
  \centering
  \includegraphics[width=0.85\linewidth]{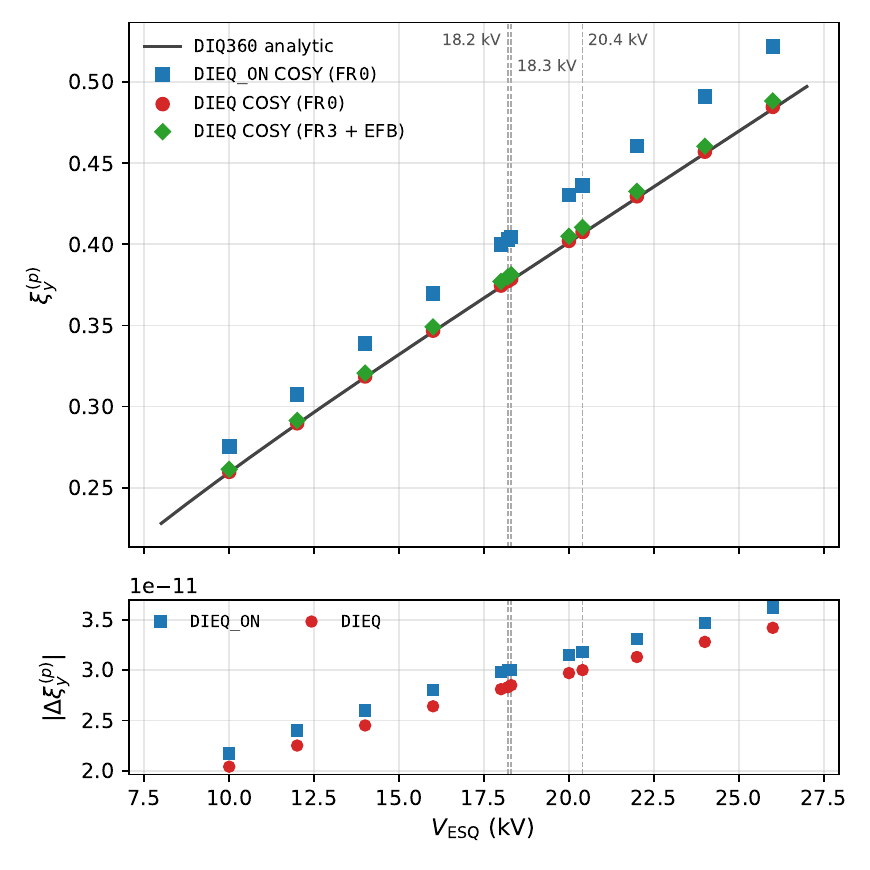}
  \caption{Vertical chromaticity $\xichromy^{(p)}$ vs ESQ voltage. Top:
    the closed-form $\mathtt{DIQ360}$ result (Eq.~\ref{eq:xiy-DIQ360},
    solid line) and COSY INFINITY DA values for the simplified modular
    $\mathtt{DIEQ\_ON}$ ring (squares, hard-edge model
    $\mathtt{FR\ 0}$), the full modular $\mathtt{DIEQ}$ ring (circles,
    $\mathtt{FR\ 0}$), and the full modular $\mathtt{DIEQ}$ ring with
    the realistic Enge fringe-field model and EFB extension (diamonds,
    $\mathtt{FR\ 3}$ + EFB). Vertical dashed gridlines mark the three
    muon $g{-}2$ storage operating points: $\Vesq = 18.2\,\mathrm{kV}$
    (Runs~3--6), $18.3\,\mathrm{kV}$ (Runs~1a/d, Run~2), and
    $20.4\,\mathrm{kV}$ (Runs~1b/c briefly). Bottom: absolute residuals
    $|\Delta\xichromy^{(p)}| = |\xiyana - \xiycosy|$ between the analytic
    chromaticity and the COSY INFINITY DA reference, for the two
    modular models in the hard-edge regime.\label{fig:chromV}}
\end{figure}

The agreement across three ring models and the $\Vesq$ sweep of
Table~\ref{tab:DIEQsweep} parallels the validation reported in
Ref.~\refcite{ChromCPO11} for the horizontal case.
Figure~\ref{fig:chromV} summarises the comparison graphically: the
closed-form $\mathtt{DIQ360}$ result tracks both modular COSY INFINITY models across
the $\Vesq$ range. The $\mathtt{DIQ360}$ continuous-ring closed-form
$\xichromy^{(p)}$ agrees with the full modular $\mathtt{DIEQ}$ values
to within ${\lesssim}\,0.3\%$ and lies ${\sim}\,6$--$7\%$ below the
$\mathtt{DIEQ\_ON}$ values across the $\Vesq$ range, a structural
difference between the ring models rather than a numerical residual;
the residual between the analytic and COSY INFINITY values within
each model is at the $3 \times 10^{-11}$ level (Table~\ref{tab:DIEQsweep}).

\textbf{Cross-check against Ref.~\refcite{ChromCPO11}.} The horizontal
chromaticity of the same modular
$\mathtt{DIEQ}$ ring (no fringe fields, Method 1C of
Ref.~\refcite{ChromCPO11}) is reproduced from the same
\texttt{sweep\_v.fox} run via $\partial \nu_x/\partial \delta_p$:
$\xichromx^{(p)}(\Vesq = 18.2\,\mathrm{kV}) = -0.12339043$, matching the
published value of Ref.~\refcite{ChromCPO11} (entry \emph{Method~1C}: $-0.1233904305126808$ analytic,
$-0.1233904305225122$ COSY INFINITY) to all eight printed digits.

\subsection{Higher-Order Chromaticities}

The chromaticities $\xi^{(j)}_{x,y}$, $j \ge 0$, are the coefficients of
the Taylor expansion of the tune in the relative momentum offset about
the on-momentum value,
\begin{equation}
  \nu_{x,y}(\delta_p) = \nu_{x,y}(0) + \sum_{j\ge 1} \xi^{(j-1)}_{x,y}\,\frac{\delta_p^j}{j!}.
  \label{eq:chrom-series}
\end{equation}
The on-momentum tune $\nu_{x,y}(0)$ is not itself a chromaticity. The
order-zero coefficients $\xi^{(0)}_{x,y} = \xichromx, \xichromy$ are the
linear chromaticities of
Eqs.~\ref{eq:xiy-DIQ360}--\ref{eq:xix-DIQ360}, and $\xi^{(j)}_{x,y}$ for
$j \ge 1$ are the higher-order (nonlinear) chromaticities. While closed-form
expressions for $\xi^{(j)}$ at $j \ge 1$ become unwieldy, the COSY INFINITY DA
framework computes them numerically up to arbitrary order. Listing~\ref{lst:chrom-o9}
shows the DA series of $\nu_x$ and $\nu_y$ up to order 9 for the full modular
$\mathtt{DIEQ}$ ring at $\Vesq = 18.2\,\mathrm{kV}$, generated by
\texttt{chrom\_o9.fox} (a single-voltage variant of
\texttt{sweep\_v.fox} with \texttt{INO~:=~9}). The order-1 coefficient is
the linear chromaticity $\xichromx, \xichromy$ already discussed; higher
orders give the nonlinear momentum dependence of the tune. The horizontal
linear coefficient $-0.1233904305$ matches Ref.~\refcite{ChromCPO11} Method~1C to all printed digits, as noted above; the corresponding
horizontal/vertical higher-order coefficients are not analysed further
in this paper, since the closed-form derivation targets the leading
$\xi^{(0)}_y$ only.

\begin{figure}[!tbp]
\begin{lstlisting}[basicstyle={\scriptsize\ttfamily},frame=shadowbox,columns=fixed,keepspaces=true]
=== DIEQ ring, V_ESQ = 18.2 kV, INO = 9, FR 0 ===
Linear tunes:  nu_x =  0.9473764793755017   nu_y =  0.3221602847213103

MU(1) (= nu_x as DA polynomial in dp):
     I  COEFFICIENT            ORDER EXPONENTS
     1  0.9473764793755017       0   0 0  0 0  0
     2  -.1233904305225120       1   0 0  0 0  1
     3  -.3681733087546968E-01   2   0 0  0 0  2
     4  -.1562867245937842E-01   3   0 0  0 0  3
     5  -.4751272809210552E-02   4   0 0  0 0  4
     6  -.7538863441833791E-02   5   0 0  0 0  5
     7  0.3282165942661314E-02   6   0 0  0 0  6
     8  -.7480427768855409E-02   7   0 0  0 0  7
     9  0.6825853021849857E-02   8   0 0  0 0  8
     -------------------------------------------

MU(2) (= nu_y as DA polynomial in dp):
     I  COEFFICIENT            ORDER EXPONENTS
     1  0.3221602847213103       0   0 0  0 0  0
     2  0.3773057590226944       1   0 0  0 0  1
     3  -.1055415252329098       2   0 0  0 0  2
     4  0.1709254868027152       3   0 0  0 0  3
     5  -.1945457652932160       4   0 0  0 0  4
     6  0.2961299492232198       5   0 0  0 0  5
     7  -.4505140119235165       6   0 0  0 0  6
     8  0.7327006441841626       7   0 0  0 0  7
     9  -1.221230140912835       8   0 0  0 0  8
     -------------------------------------------
\end{lstlisting}
\captionof{lstlisting}{DA series of the horizontal and vertical tunes for the full modular DIEQ ring at the reference voltage 18.2 kV, COSY INFINITY DA order 9, hard-edge model (\texttt{FR 0}). The order-1 coefficient is the linear chromaticity; the order-j coefficient gives the next higher-order chromaticity divided by j factorial, in the convention of the horizontal-counterpart paper.}
\label{lst:chrom-o9}
\end{figure}

\begin{figure}[tbp]
\begin{lstlisting}[basicstyle={\scriptsize\ttfamily},frame=shadowbox,columns=fixed,keepspaces=true]
=== DIEQ ring, V_ESQ = 18.2 kV, INO = 9, FR 3 + EFB ===
Linear tunes:  nu_x =  0.9468320541358946   nu_y =  0.3237823954147181

MU(1) (= nu_x as DA polynomial in dp):
     I  COEFFICIENT            ORDER EXPONENTS
     1  0.9468320541358946       0   0 0  0 0  0
     2  -.1248997001813702       1   0 0  0 0  1
     3  -.3778730608464986E-01   2   0 0  0 0  2
     4  -.1624903747627255E-01   3   0 0  0 0  3
     5  -.5219741601567675E-02   4   0 0  0 0  4
     6  -.7608998866056834E-02   5   0 0  0 0  5
     7  0.2768541237848285E-02   6   0 0  0 0  6
     8  -.7060010158752530E-02   7   0 0  0 0  7
     9  0.6203705978249155E-02   8   0 0  0 0  8
     -------------------------------------------

MU(2) (= nu_y as DA polynomial in dp):
     I  COEFFICIENT            ORDER EXPONENTS
     1  0.3237823954147181       0   0 0  0 0  0
     2  0.3798364205307833       1   0 0  0 0  1
     3  -.1046607057488083       2   0 0  0 0  2
     4  0.1406906149925861       3   0 0  0 0  3
     5   4.821595320602205       4   0 0  0 0  4
     6  -6.279906728002484       5   0 0  0 0  5
     7   12.01368125126824       6   0 0  0 0  6
     8  -19.55100816314862       7   0 0  0 0  7
     9  -3.481140292129433       8   0 0  0 0  8
     -------------------------------------------
\end{lstlisting}
\captionof{lstlisting}{DA series of the horizontal and vertical tunes for the full modular DIEQ ring at $\Vesq = 18.2$~kV, COSY INFINITY DA order 9, realistic Enge-function fringe field with EFB extension (\texttt{FR 3} + EFB; same Enge coefficients and $z_{\mathrm{EFB}} = 1.22\,\mathrm{cm}$ calibration as the horizontal-counterpart paper). Parallel to Listing~\ref{lst:chrom-o9} but with the realistic fringe-field treatment. The linear vertical chromaticity differs from the hard-edge value of Listing~\ref{lst:chrom-o9} by $\sim 0.7\%$, consistent with the prior fringe-contribution estimate from the Weisskopf dissertation integrated over the full ring.}
\label{lst:chrom-o9-fr3}
\end{figure}

\begin{figure}[t]
  \centering
  \includegraphics[width=0.78\linewidth]{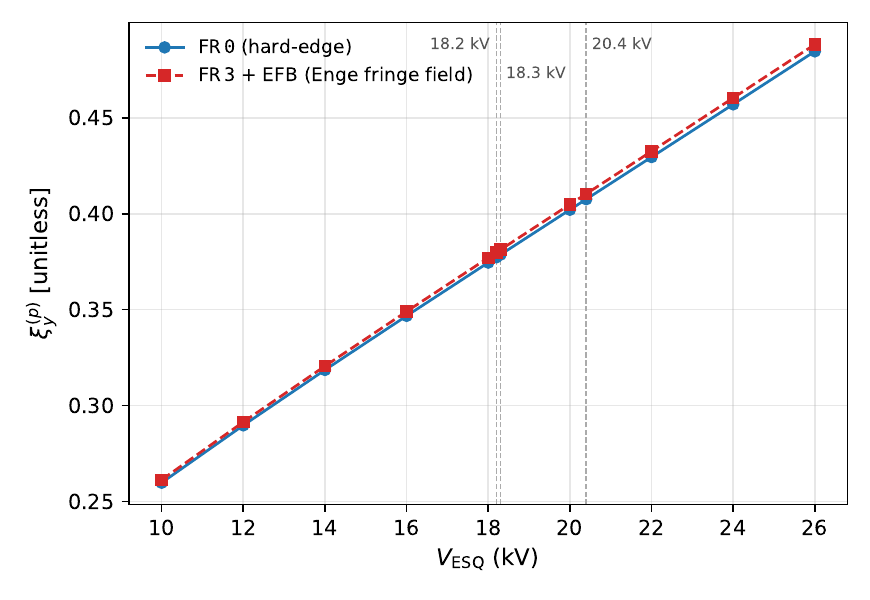}
  \caption{Linear vertical chromaticity $\xichromy^{(p)}$ of the full
    modular $\mathtt{DIEQ}$ muon $g{-}2$ ring vs ESQ voltage, COSY
    INFINITY DA: hard-edge model ($\mathtt{FR}\ \mathtt{0}$, circles,
    solid) and the realistic Enge-function fringe field with EFB
    extension ($\mathtt{FR}\ \mathtt{3}$ + EFB, squares, dashed; same
    Enge coefficients and $z_{\mathrm{EFB}} = 1.22\,\mathrm{cm}$
    calibration as the horizontal-counterpart paper\protect\cite{ChromCPO11}).
    Vertical dashed gridlines mark the storage operating voltages
    $\Vesq = 18.2\,\mathrm{kV}$ (Runs~3--6), $18.3\,\mathrm{kV}$
    (Runs~1a/d, Run~2), and $20.4\,\mathrm{kV}$ (Runs~1b/c).
    \label{fig:chromV-fringe}}
\end{figure}

The linear vertical chromaticity under the two fringe-field treatments
is compared across the full operating range in
Fig.~\ref{fig:chromV-fringe}, paralleling the linear-chromaticity
voltage scan of the horizontal-counterpart work\cite{ChromCPO11}. All
COSY INFINITY chromaticities reported here are converged in the
computational order of the DA transfer map: the hard-edge
($\mathtt{FR}\ \mathtt{0}$) linear chromaticities are already
converged at computational order~3, whereas the realistic Enge-fringe
($\mathtt{FR}\ \mathtt{3}$ + EFB) \emph{vertical} chromaticity is
converged only at order~9 and is computed at that order here
(Listing~\ref{lst:chrom-o9-fr3}; its order-3 value differs from the
converged result by ${\sim}\,4\times 10^{-2}\,\%$), while the
horizontal chromaticity is
order-3 converged in both fringe models.

The nonlinear vertical chromaticities $\xi_y^{(j)}$ reported here, in
both fringe models, are obtained with the $\mathtt{DIQ}$ pure-quadrupole
main field; in the $\mathtt{FR}\ \mathtt{3}$ + EFB case this is the
single quadrupole Enge falloff and its EFB. The allowed higher ESQ
multipoles ($12$-pole and above), which by the feed-down argument above
first contribute at $\xi_y^{(3)}$ and each carry their own effective
field boundary and Enge falloff, are not included in this series.
Incorporating them requires a per-multipole falloff extraction and a
separate Enge calibration for each harmonic in COSY INFINITY; this is
left to future work (Sec.~\ref{sec:conclusion}).

The realistic $\mathtt{FR}\ \mathtt{3}$ + EFB curve lies slightly above
the hard-edge $\mathtt{FR}\ \mathtt{0}$ curve throughout, the offset
growing smoothly with voltage; at the nominal Run~3--6 point
$\Vesq = 18.2\,\mathrm{kV}$ the two differ by ${\sim}\,0.7\%$, of the
order of the fringe-field contribution estimated previously for the
muon $g{-}2$ ring\cite{weisskopfphd}.

\section{Application to Muon $g{-}2$}
\label{sec:application}

The muon $g{-}2$ Experiment operated at three storage voltages
across Runs~1--6: $\Vesq = 18.2\,\mathrm{kV}$ (Runs~3--6 nominal,
the primary physics dataset; $\langle n \rangle \approx 0.103$),
$\Vesq = 18.3\,\mathrm{kV}$ (Runs~1a/d and Run~2;
$\langle n \rangle \approx 0.104$), and $\Vesq = 20.4\,\mathrm{kV}$
(Runs~1b and 1c, briefly;
$\langle n \rangle \approx 0.116$)\cite{g2PRAB21,weisskopfphd}.
At the Run~3--6 nominal operating point, the closed-form
result~(\ref{eq:xiy-DIQ360}) gives the vertical chromaticity
\begin{equation}
  \xichromy^{(p)} \;\approx\; 0.3765 \quad (\mathtt{DIQ360}),
  \qquad
  \xichromy^{(p)} \;\approx\; 0.3773 \quad (\text{full modular }\mathtt{DIEQ}),
  \label{eq:xiy-numerical}
\end{equation}
consistent with prior numerical results
of Ref.~\refcite{weisskopfphd}. The vertical chromaticity bears on the
pitch correction $C_p$\cite{g2PRD24,g2PRAB21}: a finite
$\xichromy$ broadens the vertical tune spread over the stored-muon momentum
distribution and thereby the average $C_p$. This analytic $\xichromy$ is not
an input to the published $C_p$ correction, which is determined from tracker
measurements of the stored-beam distribution; the closed-form result instead
serves as an independent lattice diagnostic and a tool for systematic studies
away from the nominal operating lattice. Its uncertainty is negligible on the
scale of the $78\,\mathrm{ppb}$ total systematic uncertainty of the final
muon $g{-}2$ measurement\cite{g2PRL25}.

\section{Conclusion and Outlook}
\label{sec:conclusion}

We have derived the vertical chromaticity of the Fermilab muon $g{-}2$ storage
ring in closed form for the continuous ring $\mathtt{DIQ360}$ model and via
modular composition for the $\mathtt{DIEQ\_ON}$ and full $\mathtt{DIEQ}$
ring models, using the Hamiltonian order-by-order perturbation
framework of Refs.~\refcite{AIEP108book,ESCPO10AIEP}. The closed-form
$\xichromy^{(p)}(2\pi, n)$ in Eq.~\ref{eq:xiy-DIQ360} is in direct
functional correspondence with the horizontal counterpart\cite{ChromCPO11},
exhibiting the same kinematic prefactor and a complementary $n$-dependent
geometric factor. The full set of vertical second-order aberrations of $\mathtt{DI}$ and
$\mathtt{DIQ}$ in Sec.~\ref{sec:abers}, together with the modular
composition of Sec.~\ref{sec:validation}, completes the vertical
analogue of the horizontal-plane analytic treatment\cite{ChromCPO11}
for the muon $g{-}2$ ring.

Agreement with COSY INFINITY DA at the $10^{-11}$ level across all
three ring models
confirms the analytic framework. The framework extends
naturally to the analogous derivation of the analytic $E$-field correction
$C_e$ and pitch correction $C_p$ via a DA normal-form algorithm; this,
and the extension of the reported nonlinear vertical chromaticity to the
full ESQ multipole content (with separately calibrated fringe fields per
allowed multipole, first contributing at $\xi_y^{(3)}$), is deferred to
follow-up work.

\section*{Acknowledgments}

This work was supported by the U.S.~Department of Energy under
Contract No.~DE-FG02-08ER41546 and Contract No.~DE-SC0018636. We
gratefully acknowledge our colleagues in the Beam Dynamics team of
the Fermilab Muon $g{-}2$ Experiment, where we have collaborated since
2016. This work was produced by Fermi Forward Discovery Group, LLC
under Contract No.~89243024CSC000002 with the U.S.~Department of
Energy, Office of Science, Office of High Energy Physics. Publisher
acknowledges the U.S.~Government license to provide public access
under the DOE Public Access Plan
(\mbox{https://www.energy.gov/doe-public-access-plan}).

\bibliographystyle{ws-ijmpa}
\bibliography{local_refs}

\end{document}